\documentclass[conference]{IEEEtran}
\IEEEoverridecommandlockouts
\usepackage{cite}
\usepackage{amsmath,amssymb,amsfonts,multicol}
\usepackage{algorithmic}
\usepackage{graphicx}
\usepackage{textcomp}
\usepackage[hidelinks, draft]{hyperref}
\usepackage{balance}
\usepackage{xcolor}
\usepackage{subcaption}
\usepackage{float}
\usepackage{lipsum}
\def\BibTeX{{\rm B\kern-.05em{\sc i\kern-.025em b}\kern-.08em
		T\kern-.1667em\lower.7ex\hbox{E}\kern-.125emX}}
\begin{document}
	
	\title{Impact Analysis of Antenna Array Geometry on Performance of Semi-blind Structured Channel Estimation for massive MIMO-OFDM systems\\
    	\thanks{This work has been supported by VNU University of Engineering and Technology under project number CN21.04}
	}
	
\makeatletter
\newcommand{\linebreakand}{%
\end{@IEEEauthorhalign}
\hfill\mbox{}\par
\mbox{}\hfill\begin{@IEEEauthorhalign}
}
\makeatother
	
\author{
    \IEEEauthorblockN{Do Hai Son\IEEEauthorrefmark{1}}
    \IEEEauthorblockA{dohaison1998@vnu.edu.vn}

    \and
	\IEEEauthorblockN{Tran Thi Thuy Quynh\IEEEauthorrefmark{1}}
    \IEEEauthorblockA{quynhttt@vnu.edu.vn}
    \linebreakand
    \IEEEauthorrefmark{1} VNU University of Engineering and Technology, Hanoi, Vietnam\\
}

\maketitle

\begin{abstract}
    Channel estimation is always implemented in communication systems to overcome the effect of interference and noise. Especially, in wireless communications, this task is more challenging to improve system performance while saving resources. This paper focuses on investigating the impact of geometries of antenna arrays on the performance of structured channel estimation in massive MIMO-OFDM systems. We use Cramér Rao Bound to analyze errors in two methods, i.e., training-based and semi-blind-based channel estimations. The simulation results show that the latter gets significantly better performance than the former. Besides, the system with Uniform Cylindrical Array outperforms the traditional Uniform Linear Array one in both estimation methods.
\end{abstract}


\section{Introduction}\label{Intro}
Massive MIMO (Multiple-Input Multiple-Output) is a technology in 5G wireless communications that uses a large number of antennas at the base station to communicate simultaneously with multiple user devices in the same frequency band. Massive MIMO combined with Orthogonal Frequency Division Multiplexing (OFDM) obtains the numerous benefits of coverage, capacity, spectral and energy efficiency~\cite{Elhoushy2022}. 

In communications, to recover source signals exactly, systems must get the channel state information (CSI). The known training symbols, aka pilots, are inserted in the data sequences to estimate CSI and synchronization. The length of pilot sequence should be larger than the number of elements in the antenna array. Because of the huge number of antenna elements, channel estimation in massive MIMO systems is very complex with a long training overhead~\cite{Liang2019}.

There are two popular wireless channel models that are unstructured and structured models. The unstructured model is used mostly because of simplicity but it is not really suitable for millimeter wave with several significant reflective waves. This paper relates to semi-blind channel estimation in millimeter wave MIMO-OFDM systems. Semi-blind (SB) channel estimation algorithms are combinations of conventional training-based methods and blind methods. They use several pilot symbols and other kinds of information~\cite{abed1997}. SB algorithms can reduce the number of pilot symbols efficiently but maintain acceptable accuracy~\cite{Rekik2021}. In the unstructured channel model, paths between each pair of transmitter and receiver antenna are described as complex gains~\cite{Swindlehurst2022} while the structured model includes complex gains, Directions of Departure (DoD), and Directions of Arrival (DoA). This model is also called specular or geometric channel model~\cite{Ladaycia2017, Swindlehurst2022}.

In~\cite{POORMOHAMMAD2017}, Poormohammad \textit{et al.} proposed that the geometries of antenna arrays certainly affect the accuracy of the DoA and DoD estimation. Furthermore, when the number of antennas becomes larger, 3D antenna arrays save significant installation space compared to 1D- and 2D arrays. Generally, most studies in the literature only consider a system of 1D or 2D antenna arrays. Thereby, in this work, we analysis of the performance bound of the semi-blind channel estimation in 3D-massive MIMO array geometries. For the SB method, besides the pilots part, the data symbols are assumed to be i.i.d and known statistical. The performances of systems are measured by Cramér Rao  Bound (CRB)~\cite{Ladaycia2017} for two antenna array structures, i.e., Uniform Linear Array (ULA) and Uniform Cylindrical Array (UCyA). From the simulation results, the UCyA outperforms the traditional ULA array regarding SNR and the number of elements in arrays.
 
Our contribution in this paper is to propose a CRB derivation for SB channel estimation in UCyA structures. The structured model of 3D-massive MIMO is presented in section 2. CRB deviations for training-based and SB channel estimation methods in structured and unstructured models are shown in section 3. At last, the performance of ULA and UCyA structures are compared in numerical experiments.

\section{System model}\label{SM}

This paper considers a massive MIMO-OFDM communication system in the Uplink channel with $N_t$ transmit mono-antennas and $N_r$ receive antennas with $K$ sub-carriers. 
Each OFDM symbol consists of $K$ data symbols and a CP (Cyclic Prefix) to avoid inter-symbol interference. 
At $r$-th receive antenna, after removing Cyclic Prefix and then FFT $K$-point of OFDM data samples, the output signal $\boldsymbol{y}_{r}$ in time~domain can be expressed by~\cite{Ladaycia2017}:


\begin{equation}
    \boldsymbol{y}_{r}=\sum_{j=0}^{N_{t}-1} \mathcal{F} \mathcal{T}\left({h}_{r, j}\right) \frac{\mathcal{F}}{K} \boldsymbol{x}_{j}+\boldsymbol{v}_{r}
\end{equation}
where $\mathcal{F}$ represents $K$-point discrete Fourier matrix, $\mathcal{T}$ is a circulant matrix of ${h}_{r, j}$; $\boldsymbol{x}_{j}$ is $j$-th OFDM symbol of length $K$, and $\boldsymbol{v}_{r} \in \mathbb{C}^{K \times 1}$ is an additive noise vector drawn from an i.i.d circular complex Gaussian distribution $\mathcal{C} \mathcal{N}\left(0, \sigma_{\boldsymbol{v}_r}^{2} \boldsymbol{I}_{N_{r}}\right)$.
The ${h}_{r, j}$ is an element in the vector form of full channel matrix $\boldsymbol{h} \in \mathbb{C}^{N_t N_r \times 1}$ given by:
\begin{equation}
    \label{eq:1}
        \boldsymbol{h} =\left[\boldsymbol{h}_{0}^{\top}, \boldsymbol{h}_{1}^{\top}, \ldots, \boldsymbol{h}_{N_{r} - 1 }^{\top}\right]^{\top},
        \boldsymbol{h}_{r} =\left[{h}_{r, 0}, {h}_{r, 1}, \ldots, {h}_{r, N_{t} - 1 }\right]^{\top}
\end{equation}

Assume that $L$ is the number of paths between a transmit antenna and receiver. Following the structured channel approach, we model the ${h}_{r, j}$ according to $L$ paths, complex path gains, and steering vectors, as follows:

\begin{equation}
\label{eq:2}
        h_{r, j} = \sum\limits_{l=0}^{L-1} \beta_{l, j} \cdot e^{-i k_c s(\theta_{l, j}, \phi_{l, j})} 
\end{equation}
for the $l$-th ray, $\beta$ represents complex path gain.
Zenith and azimuth angle of DoA\footnote{For simplicity, we supposed that the DoD (Direction of Departure) information is not available in the receivers.} are $\theta$, $\phi$, respectively. The $(\cdot)$ being scalar product. 
The other notations in Eq.~(\ref{eq:2}) are expressed as follows:
\begin{equation}
    \begin{aligned}
        &k_c = 2\pi/\lambda \\
        &s(\theta_{l, j}, \phi_{l, j}) = \widehat{\boldsymbol{s}} \cdot \boldsymbol{s}_p \\
        &\widehat{\boldsymbol{s}}=\sin \theta_{l, j} \cos \phi_{l, j} \widehat{\boldsymbol{x}}+\sin \theta_{l, j} \sin \phi_{l, j} \widehat{\boldsymbol{y}}+\cos \theta_{l, j} \widehat{\boldsymbol{z}} \\
        &\boldsymbol{s}_p=x_{p} \widehat{\boldsymbol{x}}+y_{p} \widehat{\boldsymbol{y}}+z_{p} \hat{\boldsymbol{z}}
    \end{aligned}
\end{equation}
where $\lambda$ is the wavelength; $\widehat{\boldsymbol{s}}$ is the unit vector in the direction of the field point; $\boldsymbol{s}_p$ is the position of $p$-th element in receiver's antenna array ($x_p, y_p, z_p$).

Particularly, we focus on two configurations of antenna arrays 
, i.e., 1D and 3D structures~\cite{POORMOHAMMAD2017}. For 1D arrays, we consider arrays with $N_{ULA}$~elements of ULA, where elements in these arrays are spaced by $d_{2D}$. For 3D arrays, UCyA consists of $N_{3D}$ layers of UCA (Uniform Circular Array) in size of $N_{UCA}$ elements. In this configuration, the spacing between two elements in UCA arrays is also $d_{2D}$ and the distance between two layers is $d_{3D}$ in $z$~direction. Thereby, the radius ($R$) of the UCA would be:
\begin{equation}
    R = \frac{1/2 \cdot d_{2D}}{\sin(\pi/N_{UCA})}
\end{equation}
The position ($\boldsymbol{s}_p$) of each element in the array structures is expressed as follows:
\begin{align} 
    &\boldsymbol{s}_p(\text{ULA}) = \;\; \begin{cases}
    x_p = n_{ULA} \times d_{2D}\\ 
    y_p = 0\\ 
    z_p = 0
    \end{cases}
    \\
    &\boldsymbol{s}_p(\text{UCyA}) = \begin{cases}
    x_p = R \times \sin(n_{UCA} \times \frac{2\pi}{N_{UCA}})\\ 
    y_p = R \times \cos(n_{UCA} \times \frac{2\pi}{N_{UCA}})\\ 
    z_p = n_{3D} \times d_{3D}
    \end{cases}
\end{align}
where $n_{ULA} = 0, 1, \ldots, N_{ULA}-1$; $n_{UCA} = 0, 1, \ldots, N_{UCA}-1$, and $n_{3D} = 0, 1, \ldots, N_{3D}-1$.

\section{CRB Derivation}\label{CRB}
In this section, we present the CRB derivations for the structured and unstructured channel models in only pilots (OP) and semi-blind (SB) estimators for 3D-massive MIMO array geometries.
\subsection{Only pilot CRB derivation}

Almost wireless communication standards use the training sequences in the physical layer (i.e., preamble) to estimate the effects of the propagation channel in the received signals. Typically, OFDM transceivers insert $K_p$ pilot symbols, which are known in both the transmitter and receiver. Thus, the receiver can exploit these pilots for channel estimation. However, there is no way to perfect accuracy in wireless communication because we cannot compute a perfect CSI. To estimate the maximum possible accuracy in wireless communication systems, CRB is used for unbiased channel estimators. Basically, the CRB is given by~\cite{Kay1993}:
\begin{equation}
    \text{CRB}(\boldsymbol{\Theta}) = \mathbf{J}_{\boldsymbol{\Theta}\boldsymbol{\Theta}}^{-1}
\end{equation}
with $\mathbf{J}_{\boldsymbol{\Theta}\boldsymbol{\Theta}}$ is the FIM (Fisher Information Matrix) and $\boldsymbol{\Theta}$ is the unknown parameters vector to be estimated. In unstructured model, $\boldsymbol{\Theta} \simeq	 \boldsymbol{h}$~\cite{Ladaycia2017}, FIM is associated to the known pilots denoted by $\mathbf{J}_{\boldsymbol{\Theta}\boldsymbol{\Theta}}^p$. Therefore, the parameters vector to be estimated is expressed by~\cite{Menni2012}\footnote{We ignored noise powers ($\sigma^2_{\boldsymbol{v}}$) since its estimation error does not affect the desired $\boldsymbol{h}$.}:
\begin{equation}
    \boldsymbol{\Theta}=\left[\boldsymbol{h}^{\top},  \quad  \left(\boldsymbol{h}^{*}\right)^{\top}\right]
\end{equation}

In massive MIMO-OFDM systems, $K_p$ pilots are arranged in OFDM symbols~\cite{Garro2020}, and since the noise is an i.i.d random process, we could formulate FIM in the OP case as follows:
\begin{equation}
\label{eq:9}
    \mathbf{J}_{\boldsymbol{\Theta} \boldsymbol{\Theta}}^{p}=\sum_{i=1}^{K_{p}} \mathbf{J}_{\boldsymbol{\Theta} \boldsymbol{\Theta}}^{p_{i}}
\end{equation}
with $\mathbf{J}_{\boldsymbol{\Theta} \boldsymbol{\Theta}}^{p_{i}}$ is the FIM associated with the $i$-th pilot~\cite{Kay1993} given by:

\begin{equation}
    \label{eq:10}
    \begin{aligned}
        \mathbf{J}_{\boldsymbol{\Theta} \boldsymbol{\Theta}}^{p_{i}} &=\mathbb{E}\left\{\left(\frac{\partial \ln p(\mathbf{y}(i), \boldsymbol{h})}{\partial \boldsymbol{\Theta}^{*}}\right)\left(\frac{\partial \ln p(\mathbf{y}(i), \boldsymbol{h})}{\partial \boldsymbol{\Theta}^{*}}\right)^{H}\right\} \\
    \end{aligned}
\end{equation}
where $\mathbb{E}$ is the expectation operator; $p(\mathbf{y}(i), \boldsymbol{h})$ is the probability density function (pdf) of the received signal given $\boldsymbol{h}$. Eq.~(\ref{eq:10}) is complex derivations. Hence, it can be expressed by:
\begin{equation}
    \mathbf{J}_{\boldsymbol{\theta} \boldsymbol{\theta}}^{p_{i}}=\frac{\boldsymbol{x}(i)^{H} \boldsymbol{x}(i)}{\sigma_{\boldsymbol{v}}^{2}}
\end{equation}

When considering a structured model as shown in~(\ref{eq:2}), the parameters vector of size \mbox{$4N_t~\times L$} to be estimated is given by:
\begin{equation}
    \boldsymbol{\Theta}=\left[ \boldsymbol{\beta}^\top, \quad \boldsymbol{(\beta^*)}^\top, \quad \boldsymbol{\theta}^\top, \quad \boldsymbol{\phi}^\top \right]
\end{equation}
with the complex gain, the conjugate of complex gain, zenith, and azimuth angle of DoA vectors of size $N_t~\times~L$ respectively are $\boldsymbol{\beta}=\left[\beta_{0,0}, \ldots, \beta_{L-1, N_t -1}\right]^{\top}$, $\boldsymbol{\beta^*}=\left[\beta^*_{0,0}, \ldots, \beta^*_{L-1, N_t - 1}\right]^{\top}$, $\boldsymbol{\theta}=\left[\theta_{0,0}, \ldots, \theta_{L-1, N_t - 1}\right]^{\top}$, and $\boldsymbol{\phi}=\left[\phi_{0,0}, \ldots, \phi_{L-1, N_t - 1}\right]^{\top}$. Regarding to the FIM derivation of parameters transformation~\cite{Kay1993}, the FIM ($\mathbf{J}^p_{\boldsymbol{h} \boldsymbol{h}}$) of channel $\boldsymbol{h}$ in (\ref{eq:1}) would be:

\begin{equation}
\label{eq:13}
    \mathbf{J}^p_{\boldsymbol{h} \boldsymbol{h}}=\frac{\partial \boldsymbol{h}}{\partial \boldsymbol{\Theta}} \mathbf{J}^p_{\Theta \Theta} {\frac{\partial \boldsymbol{h}}{\partial \boldsymbol{\Theta}}}^{H}
\end{equation}
where 
\begin{equation}
    \frac{\partial \boldsymbol{h}}{\partial \boldsymbol{\Theta}}=
    \left[\frac{\partial \boldsymbol{h}}{\partial \boldsymbol{\beta}}, 
    \frac{\partial \boldsymbol{h}}{\partial \boldsymbol{\beta^*}},
    \frac{\partial \boldsymbol{h}}{\partial \boldsymbol{\theta}}, 
    \frac{\partial \boldsymbol{h}}{\partial \boldsymbol{\phi}}\right]
\end{equation}
More particularly, we express the derivations as follows:
\begin{subequations}
    \begin{align}
    &\frac{\partial \boldsymbol{h}}{\partial \boldsymbol{\beta}}=
    \left[\begin{array}{llll}
        \boldsymbol{B}_{0}^{\top}, & \boldsymbol{B}_{1}^{\top}, & \ldots, & \boldsymbol{B}_{N_{r} - 1}^{\top}
    \end{array}\right]^{\top}\\
    &\boldsymbol{B}_{r}=\operatorname{diag}\left(\left[\boldsymbol{B}_{r, 0}, \quad \boldsymbol{B}_{r, 1}, \quad \ldots, \quad \boldsymbol{B}_{r, N_{t} - 1}\right]\right) \\
    &\boldsymbol{B}_{r, j}=\left[\begin{array}{cccc}
        \frac{\partial h_{r, j}}{\partial \beta_{0, j}} &
        \frac{\partial h_{r, j}}{\partial \beta_{1, j}} & 
        \ldots & 
        \frac{\partial h_{r, j}}{\partial \beta_{L-1, j}}
    \end{array}\right]^\top
    \end{align}
\end{subequations}
where the derivations are precisely provided in (\ref{eq:15}).

\begin{figure*}[]
    \begin{subequations}
    \label{eq:15}
        \begin{flalign}
        &\frac{\partial {h}_{r, j}}{\partial \beta_{l, j}}= \frac{1}{2} (1 - i) \cdot e^{-i k_c s(\theta_{l, j}, \phi_{l, j})} &  \\
        &\frac{\partial {h}_{r, j}}{\partial \beta^*_{l, j}}= \frac{1}{2} (1 + i) \cdot e^{-i k_c s(\theta_{l, j}, \phi_{l, j})} \\
        &\frac{\partial {h}_{r, j}}{\partial \theta_{l, j}}=
        \beta_{l, j} 
        [-i k_c (\cos\theta_{l, j} \cos\phi_{l, j} x_p + \cos\theta_{l, j} \sin\phi_{l, j} y_p
        - \sin \theta_{l, j} z_p)] \cdot e^{-j k_c s(\theta_{l, j}, \phi_{l, j})} \\ 
        &\frac{\partial {h}_{r, j}}{\partial \phi_{l, j}}=\beta_{l, j}
         [-i k_c (-\sin\theta_{l, j} \sin\phi_{l, j} x_p + \sin\theta_{l, j} \cos\phi_{l, j} \cdot y_p 
        + \cos \theta_{l, j} z_p)] \cdot e^{-i k_c s(\theta_{l, j}, \phi_{l, j})} 
        \end{flalign}
        \hfill
    \end{subequations}
    \hrule
\end{figure*}

\subsection{Semi-blind CRB derivation}

In the SB approach, besides using pilots, estimators also use other information of the unknown data to aid in channel estimation. In this paper, we supposed that pilots and data in OFDM symbols are statistically independent. So, the FIM of this strategy is formulated as follows:
\begin{equation}
    \label{eq:17}
    \mathbf{J}_{\boldsymbol{\Theta} \boldsymbol{\Theta}}^{SB}= \mathbf{J}_{\boldsymbol{\Theta} \boldsymbol{\Theta}}^{p} + \mathbf{J}_{\boldsymbol{\Theta} \boldsymbol{\Theta}}^{d}
\end{equation}
where $\mathbf{J}_{\boldsymbol{\Theta} \boldsymbol{\Theta}}^{d}$ is FIM associated with the unknown data while $\mathbf{J}_{\boldsymbol{\Theta} \boldsymbol{\Theta}}^{p}$ is related to the known pilots as formulated in Eq.~(\ref{eq:9}). The $K_d$ unknown data are assumed to be i.i.d, with zero mean and a covariance matrix $\mathbf{C}_{\boldsymbol{x}}=\operatorname{diag}\left(\boldsymbol{\sigma}_{\boldsymbol{x}}^{2}\right)$ where $\boldsymbol{\sigma}_{\boldsymbol{x}}^{2} \stackrel{\operatorname{def}}{=}\left[\sigma_{\boldsymbol{x}_{0}}^{2}, \ldots, \sigma_{\boldsymbol{x}_{N_{t}-1}}^{2}\right]^\top$ with $\sigma_{\boldsymbol{x}_i^{2}}$ is the transmit power of $i$-th transmit antenna. The covariance matrix $\mathbf{C}_{\boldsymbol{y}}$ becomes:

\begin{equation}
    \mathbf{C}_{\boldsymbol{y}}=\sum_{i=0}^{N_{t}-1} \sigma_{\boldsymbol{x}_{i}}^{2} \boldsymbol{\lambda}_{i} \boldsymbol{\lambda}_{i}^{H}+\sigma_{\boldsymbol{v}}^{2} \mathbf{I}_{K N_{r}}
\end{equation}
where $\mathbf{I}_{K N_{r}}$ is the identify matrix of size $K N_r$ and $\boldsymbol{\lambda}$ is defined as:
$$
\boldsymbol{\lambda}=\left[\boldsymbol{\lambda}_{0}, \boldsymbol{\lambda}_{1}, \ldots, \boldsymbol{\lambda}_{N_{t}-1}\right], \quad \boldsymbol{\lambda}_{j}=\left[\boldsymbol{\lambda}_{0, j}, \boldsymbol{\lambda}_{1, j}, \ldots, \boldsymbol{\lambda}_{N_{r}-1, j}\right]^{\top}
$$
with $\boldsymbol{\lambda}_{r, j}=\operatorname{diag}\left(\mathcal{F}_0 h_{r, j}\right)$ and $\mathcal{F}_0$ is the first column of matrix $\mathcal{F}$. The FIM of data has the following form:
\begin{equation}
    \mathbf{J}_{\boldsymbol{\Theta} \boldsymbol{\Theta}}^{d}=K_{d}\left[\begin{array}{cc}
    \mathbf{J}_{\boldsymbol{h} \boldsymbol{h}}^{d} & \mathbf{J}_{\boldsymbol{h} \boldsymbol{h}^{*}}^{d} \\
    \mathbf{J}_{\boldsymbol{h}^{\star} \boldsymbol{h}}^{d} & \mathbf{J}_{\boldsymbol{h}^{\star} \boldsymbol{h}^{*}}^{d}
    \end{array}\right]
\end{equation}

The FIM $\mathbf{J}_{\boldsymbol{\Theta} \boldsymbol{\Theta}}^{d}$ of data is given by~\cite{Kay1993}:
\begin{equation}
    \mathbf{J}_{\boldsymbol{\Theta} \boldsymbol{\Theta}}^{d}=\operatorname{tr}\left\{\mathbf{C}_{\boldsymbol{y}}^{-1} \frac{\partial \mathbf{C}_{\boldsymbol{y}}}{\partial \boldsymbol{h}^{*}} \mathbf{C}_{\boldsymbol{y}}^{-1}\left(\frac{\partial \mathbf{C}_{\boldsymbol{y}}}{\partial \boldsymbol{h}^{*}}\right)^{H}\right\}
\end{equation}
with $\frac{\partial \mathbf{C}_{\boldsymbol{y}}}{\partial \boldsymbol{h}_{i}^{*}}=\boldsymbol{\lambda} \mathbf{C}_{\boldsymbol{x}} \frac{\partial \boldsymbol{\lambda}^{H}}{\partial \boldsymbol{h}_{i}^{*}}$. In the unstructured channel model approach, the CRB of the SB method is the inverse of (\ref{eq:17}). Similar to the OP method, the CRB of the SB method using the structured channel model is given by applying (\ref{eq:17}) to the transformation in (\ref{eq:13}).

\section{Simulation Results}\label{SR}


In this section, we simulate three scenarios to verify the performance of SB and UCyA antenna structures. In detail, there are channel estimation CRBs versus, i.e., SNR (signal noise ratio), the number of UCyA layers $N_{3D}$, and the number of UCA elements $N_{UCA}$. The simulation parameters of a massive MIMO-OFDM system are shown in Table~\ref{tab:simulation_param}~\cite{Swindlehurst2022}. The results are obtained by averaging 1,000 running times.
\begin{table}[H]
\centering
\caption{Simulation parameters}
\label{tab:simulation_param}
\begin{tabular}{p{3.5cm} | p{4cm}}
\hline
\hline
\textbf{Parameters} & \textbf{Specifications} \\ \hline
Number of transmit antennas                            & $N_t = 2$      \\ \hline
Antenna spacing                 & $d_{2D} = d_{3D} = \lambda / 2$ \\ \hline
Number of paths                   & $L = 4$      \\ \hline
Sub-carriers                    & $K = 64$     \\ \hline
Pilot, Data symbols                          & $K_p = 16, K_d = 48$     \\ \hline
Complex path gain               & $\beta \sim \mathcal{C} \mathcal{N}\left(0, 1 \right)$     \\ \hline
Azimuth angle of DoA            & $\phi^\circ \sim \mathcal{U}(-\pi/2, \pi/2)$        \\ \hline
Zenith angle of DoA           & $\theta^\circ \sim \mathcal{U}(-\pi/2, \pi/2)$       \\ \hline
\end{tabular}
\end{table}

In Fig.~\ref{fig:op}, the number receive antennas is 96 where $N_{ULA}~=~96, N_{UCA} = 24,$ and $N_{3D} = 4$. Overall, CRB curves of the structure channel model approach clearly outperform those of the unstructured at $10^3$~dB of gain. For (SNR $\le 5$~dB), the unstructured CRB of SB method is slightly better than the OP one. In the structured approach, the difference between the CRB of OP and SB methods is evident when (SNR $\le 5$~dB) and remains stable at higher SNR values. In massive MIMO array geometries, the CRBs of UCyA structures are higher accurate than those of ULA structures in both OP and SB methods. Thus, it can be shown that using structured models, SB estimation methods, and 3D antenna arrays can provide higher performance for channel estimators in massive MIMO systems.
\begin{figure}[t]
    \centering
    \includegraphics[width=\linewidth]{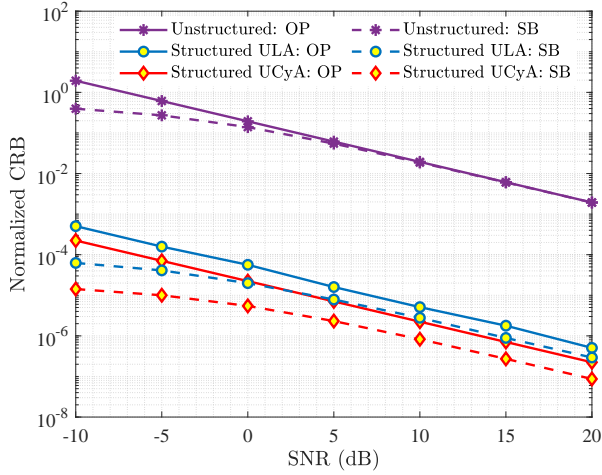}
    \caption{CRB for ULA and UCyA vs. structured and unstructured approaches. Configurations of antenna arrays are $N_{ULA} = 96, N_{UCA} = 24, N_{3D} = 4$.}
    \label{fig:op}
\end{figure}

In Fig.~\ref{fig:op_N3D}, the number of layers $N_{3D}$ in the UCyA structures is investigated by fixing $N_{UCA}$ elements at $24$ and SNR~$=5$~dB. Again, the CRBs of the structured channel model give superior quality to those of the unstructured model. The first point, when the number of antennas increases, the CRB of the unstructured model also increases. Moreover, the SB method is also almost trivial in this case. On the other hand, CRBs in the structured approach tend to decrease as the number of antennas increases until all of them converge to a point at $10^{-6}$. At $N_{3D}$ values as low as 2 to 6 layers, UCyA structures give a relatively significant quality when compared to the ULA in both NB and SB methods. Hence, the structured model method gives a better channel estimation error rate, while 3D antenna arrays are valuable when the number of layers is small. However, note that, in addition to the advantage of accuracy in channel estimation, UCyA structures save powerful deployment areas.
\begin{figure}[t]
    \centering
    \includegraphics[width=\linewidth]{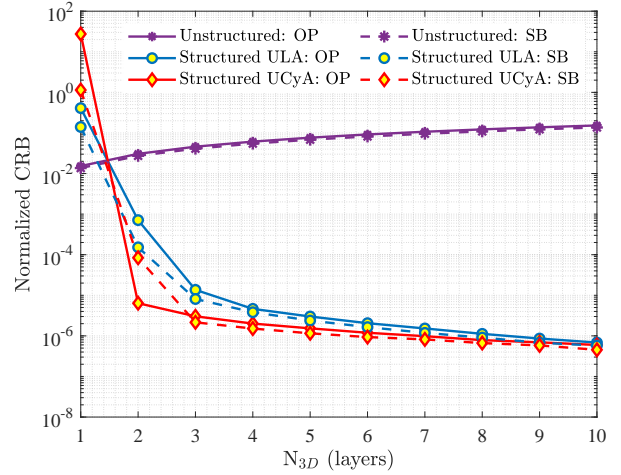}
    \caption{CRB for ULA and UCyA vs. number of $N_{3D}$. The simulation parameters are $N_{UCA} = 24, N_{ULA} = 24 * N_{3D}$, and SNR $=5$~dB.}
    \label{fig:op_N3D}
\end{figure}
\begin{figure}[H]
    \centering
    \includegraphics[width=\linewidth]{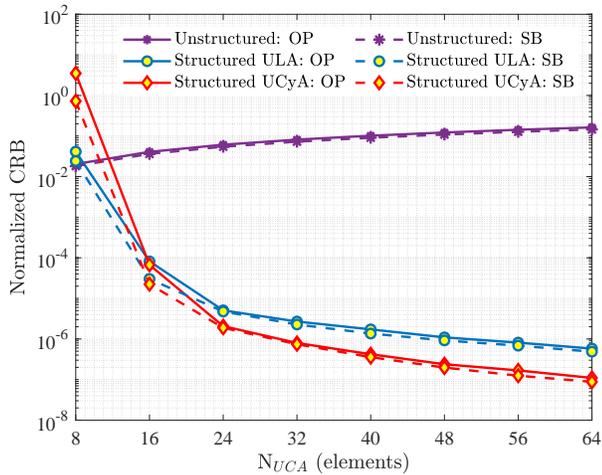}
    \caption{CRB for ULA and UCyA vs. number of $N_{UCA}$. The simulation parameters are $N_{3D} = 4, N_{ULA} = 4 * N_{UCA}$, and SNR $=5$~dB.}
    \label{fig:op_NUCA}
\end{figure}

At last, in Fig.~\ref{fig:op_NUCA} the number of UCA elements $N_{UCA}$ is turn in range 8 to 64 elements while $N_{3D} = 4, N_{ULA} = 4 * N_{UCA}$, and SNR $=5$~dB. Since the unstructured performance is only affected by the number of antennas, its performance remains at the same levels as shown in Fig.~\ref{fig:op_N3D}. However, instead of converging as before in Fig.~\ref{fig:op_N3D}, the CRBs in the structured approach gradually reduce as $N_{UCA}$ continues to rise. At large $N_{UCA}$ elements, UCyA arrays linearly perform better than ULA arrays. Note that, despite the accuracy benefits, it is more complicated to produce large UCA arrays.

\section{Conclusion}
This paper uses Cramér Rao Bound to analyze the effect of antenna array geometry on channel estimation errors in massive MIMO-OFDM systems. The CRBs of channel estimation in both cases, i.e., OP and SB-based methods, are presented. The simulation results demonstrate that the structured channel model significantly improves the channel estimation performance. The UCyA structure obtains fewer channel estimation errors in this model, and this geometry is more suitable than the traditional ULA structure. 

\section*{Dedication}

This paper is part of a special session at SSP 2023 that honors Professor Huynh Huu Tue, a distinguished academic in the field of signal processing and a proud Vietnamese Canadian.
	
\bibliographystyle{IEEEtran}
\bibliography{library.bib}	
	
\end{document}